\documentclass[%
article,
twocolumn,
superscriptaddress,
amsmath,amssymb,
aps,
journal=prb,
]{revtex4-2}

\usepackage{natbib}
\usepackage{graphicx}
\usepackage{dcolumn}
\usepackage{bm}
\usepackage{xcolor}
\usepackage{float}
\usepackage{newtxtext,newtxmath}
\usepackage{cleveref}

\renewcommand{\Re}{\mathrm{Re}}
\renewcommand{\Im}{\mathrm{Im}}

\begin{document}

\title{
Strong anharmonicity dictates ultralow thermal conductivities of type-I clathrates
}

\author{Dipti Jasrasaria}
\email{dj2667@columbia.edu}
\affiliation{Department of Chemistry, Columbia University, New York, New York 10027, USA}

\author{Timothy C. Berkelbach}
\email{t.berkelbach@columbia.edu}
\affiliation{Department of Chemistry, Columbia University, New York, New York 10027, USA}
\affiliation{Initiative for Computational Catalysis, Flatiron Institute, New York, New York 10010, USA}

\date{\today}

\begin{abstract}
Type-I clathrate solids have attracted significant interest due to their ultralow thermal conductivities and subsequent promise for thermoelectric applications, yet the mechanisms underlying these properties are not well understood. Here, we extend the framework of vibrational dynamical mean-field theory (VDMFT) to calculate temperature-dependent thermal transport properties of solids using a many-body Green's function approach. When applied to a coarse-grained description of $X_8$Ga$_{16}$Ge$_{30}$, where $X=$ Ba, Sr, we find that nonresonant scattering between cage acoustic modes and rattling modes leads to a reduction of acoustic phonon lifetimes and thus thermal conductivities. Moreover, we find that the moderate temperature dependence of conductivities above 300\,K, which is consistent with experimental measurements, cannot be reproduced by textbook perturbation theory calculations, which predict a $T^{-1}$ dependence. Therefore, we suggest that nonperturbative anharmonic effects, including four- and higher-phonon scattering processes, are responsible for the ultralow thermal conductivities of type-I clathrates.

\end{abstract}

\maketitle

\section{Introduction}

Good thermoelectric materials, which have the potential to convert waste heat to useful electricity~\cite{zhang2015thermoelectric, zhang2014organic, fergus2012oxide}, have high electronic conductivities but low thermal conductivities,
which presents a challenging problem in materials design~\cite{shi_advanced_2020}. 
One promising class of materials are type-I clathrate solids, which contain
cage-like chemical structures of covalently bonded atoms that host loosely
bound ``guest'' atoms (Fig.~\ref{fig1:model-vdmft}a), although the mechanism responsible for their low thermal
conductivity has been debated~\cite{dong_theoretical_2001, christensen_avoided_2008, tadano_impact_2015, tadano_quartic_2018, ikeda_kondo-like_2019, lindroth_thermal_2019,  roy_occupational_2023}.
The rattling displacements of the guest atoms generate dispersionless, low-frequency optical
phonon branches that intersect the acoustic phonon branches of the cage lattice~\cite{christensen_avoided_2008}.
The hybridization of these branches suppresses the group velocity
of the acoustic phonons, which also inherit the reduced lifetimes of the strongly anharmonic
optical rattling phonons~\cite{tadano_impact_2015, jasrasaria2024}.
More exotic behaviors have also been proposed, including a phonon Kondo effect~\cite{ikeda_kondo-like_2019}.

One of the simplest microscopic approaches to calculate the thermal conductivity tensor, $\kappa^{ij}$,
is based on the Boltzmann transport equation in the single-mode approximation (BTE-SMA)~\cite{ziman2001electrons}, 
\begin{equation}
\label{eq:kappa_bte-sma}
    \kappa^{ij}_{\text{SMA}} = \frac{1}{V}\sum_{\bm{k}} \sum_\lambda C_\lambda(\bm{k})v^i_\lambda(\bm{k})v^j_\lambda(\bm{k})\tau_\lambda(\bm{k})\,,
\end{equation}
where $V$ is the cell volume, $\lambda$ is an index over phonon modes in the system,
$C_\lambda(\bm{k})$ is the heat capacity of mode $\lambda$, and
$v^i_\lambda(\bm{k})$ is the phonon group velocity of mode $\lambda$ in Cartesian direction $i$.
%
The mode lifetimes, $\tau_\lambda(\bm{k})$, are commonly
obtained by treating three-phonon processes to second order in perturbation theory (PT).
Given the
strong anharmonicity required to suppress the lifetimes of acoustic phonons in
materials with low thermal conductivities, it is natural to question the
quantitative accuracy of this approach.
In particular, at temperatures above the Debye temperature, the heat capacity is constant,
and PT predicts that the lifetimes decrease as $\tau \propto T^{-1}$, such that $\kappa \propto T^{-1}$.
Experimentally observed deviations from this temperature dependence are indicative of stronger anharmonicity~\cite{sales_structural_2001, may_characterization_2009},
and their theoretical explanation requires more advanced methods. For example, four-phonon scattering and temperature-dependent phonon renormalization can be accounted for using higher-order PT~\cite{Feng_PhysRevB_2017, Ravichandran_PhysRevB_2018, Xia_PhysRevX_2020} or self-consistent phonon theory~\cite{Hellman_PhysRevB2014, tadano_quartic_2018, Xia_PhysRevX_2020}. Additionally, recent approaches based on Wigner transport and Green-Kubo linear response theory~\cite{simoncelli_unified_2019, simoncelli2022wigner, isaeva_modeling_2019} are able to describe effects, such as interband transport (further discussed in Sec.~II.C), that are neglected in the BTE-SMA. These methods have been shown to successfully capture deviations from the $T^{-1}$ behavior in various materials with low thermal conductivities.

Nonperturbative methods are not reliant on finite-order PT and therefore inherently include all orders of phonon scattering processes. Approaches based on equilibrium and nonequilibrium molecular dynamics (MD) simulations~\cite{dong_theoretical_2001, schelling_comparison_2002, tretiakov_thermal_2004, english2017equilibrium, carbogno_ab_2017} include nonperturbative effects of strong anharmonic interactions, but the computational expense associated with such direct simulation, especially when considering finite-size effects, makes them intractable for systems with complex unit cells. Furthermore, obtaining systematic or mechanistic insight from such large-scale atomistic simulations can be challenging relative to approaches, such as the BTE-SMA, which calculate phonon-mode resolved thermal conductivities.
Recently, we introduced vibrational dynamical mean-field theory (VDMFT) as an affordable
but nonperturbative computational method for simulating anharmonic lattice 
dynamics~\cite{shih_anharmonic_2022}, and we used it to calculate the phonon spectral function
of the type-I clathrates Ba$_8$Ga$_{16}$Ge$_{30}$ (BaGG) and Sr$_8$Ga$_{16}$Ge$_{30}$ (SrGG)~\cite{jasrasaria2024}.


The present work has two goals. The first goal, which is methodological, is to extend VDMFT to enable the calculation of thermal conductivities. Because DMFT is a theory of the one-body Green's function, it does not formally predict two-body Green's functions needed to calculate transport coefficients. In electronic structure theory, where DMFT is a mature method, this limitation is commonly addressed by approximating the two-body Green's function in terms of the one-body Green's function (i.e., neglecting vertex corrections \cite{GeorgesRevModPhys1996, HaulePRB2010}), yielding electronic resistivities that exhibit signatures of strong electron correlations and good agreement with experiments \cite{merino_transport_2000, Deng_PRL2016, redka2024interplay}. Here, we provide the first implementation and test of such a framework for lattice thermal conductivities.
We choose to test these methodological developments on model  systems that are simple enough to be studied in detail with the present theory but complex enough to be representative of real materials of interest. This defines our second goal, which is to provide insight into the possible roles of strong anharmonicity on the temperature-dependent transport properties of type-I clathrate materials.

Through our VDMFT calculations on model clathrate systems, we confirm that anharmonic interactions between acoustic modes of the cage lattice and flat optical modes corresponding to rattling motions of guest atoms inside 24-atom cages are responsible for the ultralow thermal conductivities of these materials. Furthermore, through a nonperturbative description of this anharmonicity, we find that the thermal conductivity has a relatively moderate dependence on the temperature above 300\,K, consistent with experimental observations. Within our model, these results cannot be captured, even qualitatively, using conventional lowest-order PT, suggesting that nonperturbative effects of anharmonicity, including four-phonon and higher-order scattering processes, are important for understanding the reduced thermal transport of type-I clathrates.
We also briefly evaluate the evidence for a phonon Kondo effect~\cite{ikeda_kondo-like_2019}, which our method is uniquely suited to study, given the success of DMFT and its associated impurity problem in explaining Kondo physics in electronic systems.

\section{Methods}

\subsection{Clathrate model}

To study the vibrational structure and thermal transport properties of type-I clathrates, we use a coarse-grained model of the material~\cite{jasrasaria2024}, illustrated in Fig.~\ref{fig1:model-vdmft}b. We model dodecahedral and tetrakaidecahedral cages with single ``hollow" atoms, which are arranged on an FCC lattice and interact through a Lennard-Jones (LJ) potential. While the model neglects high-frequency intracage dynamics, the resulting modes have small group velocities and high frequencies and will therefore have minor contributions to the overall thermal conductivity~\cite{tadano_impact_2015}.
Guest atoms are described by smaller atoms at each FCC site, which interact with the cage atoms at those sites through anharmonic, quartic potentials. The Hamiltonian for our clathrate model is given by
\begin{gather}
\begin{split}
H &= \frac{1}{2}\sum_{\bm{m}\alpha}\left(\frac{\bm{p}_{\bm{m}\alpha}^{2}}{m_{\alpha}}\right)+\frac{1}{2}\sum_{\bm{m}\alpha,\bm{n}\beta}^{\text{cage}} V_{\text{LJ}}\big(\left|\bm{r}_{\bm{m}\alpha}-\bm{r}_{\bm{n}\beta}\big|\right) \\
&\hspace{1em} +\sum_{\bm{m}}\sum_{\alpha}^{\text{cage}}\sum_{\beta}^{\text{guest}}{}^{'} V_{\text{q}}^{\alpha\beta}\left(\bm{r}_{\bm{m}\alpha}-\bm{r}_{\bm{m}\beta}\right)\label{eq:Hamiltonian}
\end{split}\\
V_{\text{LJ}}\left(r\right) = 4\epsilon\left[\left(\frac{\sigma}{r}\right)^{12}-\left(\frac{\sigma}{r}\right)^{6}\right]\label{eq:V_LJ}\\
V_{\text{q}}^{\alpha\beta}\left(\bm{r}\right) = \sum_{i}\left(\frac{1}{2}K_{\beta,i}r_{i}^{2}+g_{\beta,i}r_{i}^{4}\right)\,,\label{eq:V_q}
\end{gather}
where the primed summation indicates that only cages and guests on the same lattice site interact. Here, $\bm{m},\bm{n}$ are lattice translation vectors, $\alpha,\beta$ are indices over atoms in the unit cell, and $i$ is an index over the Cartesian directions. The position, momentum, and mass of atom $\alpha$ in cell $\bm{m}$ are given by $\bm{r}_{\bm{m}\alpha}$, $\bm{p}_{\bm{m}\alpha}$, and $m_{\alpha}$, respectively. Further details about our model, including parameters for the LJ and quartic potentials, are given in the Supplemental Material (SM)~\footnote{See Supplemental Material at [XXX] for details about clathrate model and parameters; details of molecular dynamics, self-consistent phonon, and perturbation theory calculations; figures regarding phonon mean-free-paths and effects of grain boundary scattering; figures comparing thermal conductivities calculated using different methods; and analysis of the temperature dependence of thermal conductivities. The Supplemental Material also includes Refs.~[\onlinecite{plimpton1995fast, Parlinski1997, Zhou2014}]}\nocite{plimpton1995fast, Parlinski1997, Zhou2014}. 
Although our model is parameterized against first-principles calculations~\cite{jasrasaria2024}, we note that it has vanishing cubic force constants involving guest atoms, i.e., $\partial^3 \mathcal{V}/\partial u_{\bm{m}\alpha i} \partial u_{\bm{n}\beta j} \partial u_{\bm{l} \gamma k} = 0$, where $\alpha,\beta$ and/or $\gamma$ are guest atoms, although this is not true of the fully atomistic material. This difference has important implications for the performance of PT and comparison to previous works, as discussed below.

\subsection{Anharmonic lattice dynamics with vibrational dynamical mean-field theory}

\begin{figure*}[ht]
\includegraphics[width=6.75in]{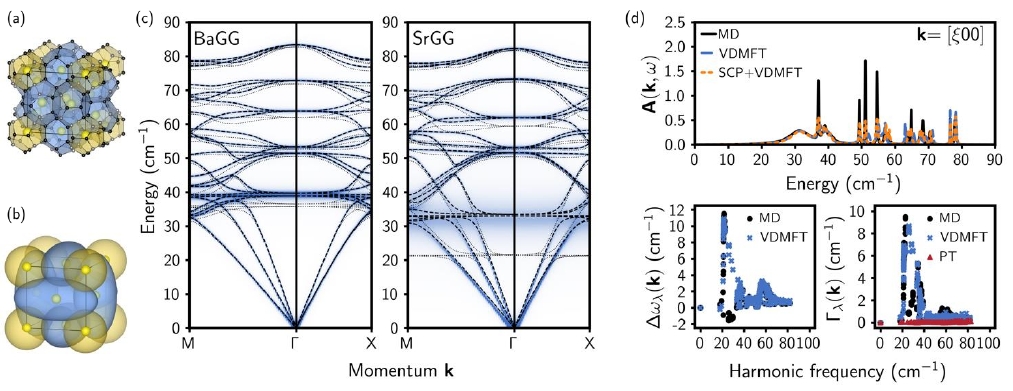}
\caption{
(a) Crystal structure of BaGG \cite{bentien_crystal_2005}, where Ba(1) atoms are in dodecahedral cages (gold) and Ba(2) atoms are in tetrakaidecahedral cages (blue). (b) Schematic of coarse-grained model of filled clathrates BaGG and SrGG, where the large cage atoms are colored according to their quartic cage-guest potentials. (c) Spectral functions of BaGG (left) and SrGG (right) at 300\,K calculated using SCP+VDMFT. Black dotted lines indicate the harmonic dispersion relation, and black dashed lines indicate the renormalized dispersion calculated using SCP. (d) The spectral function of SrGG at 300\,K at $\bm{k}=[\xi 0 0 ]$, where $\xi=\pi/a$, calculated using MD, VDMFT, and SCP+VDMFT (top). Anharmonic frequency shifts (bottom left) and linewidths (bottom right) of SrGG at 300\,K obtained from the self-energy calculated using MD, SCP+VDMFT, and SCP+PT sampled on a $4\times 4 \times 4$ Gamma-centered grid of the BZ. The frequency shift is calculated as $\Delta \omega_\lambda (\bm{k}) = \omega_{\text{eff},\lambda}(\bm{k}) - \omega_\lambda(\bm{k})$, where $\omega^2_{\text{eff},\lambda}(\bm{k}) = \omega^2_\lambda(\bm{k}) + 2\omega_\lambda(\bm{k}) \Re\bm{\pi}_{\lambda,\lambda} \big(\bm{k}, \omega_{\text{eff},\lambda}(\bm{k})\big)$ and was solved for iteratively. The linewidth is calculated as $\Gamma_\lambda (\bm{k}) = 2\omega_\lambda(\bm{k})\Im \bm{\pi}_{\lambda,\lambda} \big(\bm{k}, \omega_{\text{eff},\lambda} (\bm{k})\big) / \omega_{\text{eff}, \lambda} (\bm{k})$.}
\label{fig1:model-vdmft}
\end{figure*}

In VDMFT, we calculate the anharmonic phonon Green's function (GF) of the periodic lattice~\cite{Cowley1963},
\begin{equation}
    D_{\lambda,\lambda'}(\bm{k},t) = -i\theta(t) \langle \left[u_\lambda(\bm{k},t), u_{\lambda'}(-\bm{k},0)\right] \rangle\,,
\end{equation}
where 
$\langle \cdot \rangle$ denotes an equilibrium average at temperature $T$.
Here, $u_\lambda(\bm{k})$ is the phonon mode displacement,
\begin{equation}
    u_\lambda(\bm{k}) = \sum_{\alpha i} c_{\alpha i, \lambda}(\bm{k}) u_{\alpha i}(\bm{k})\,,
\end{equation}
expressed in the basis of translational symmetry adapted, mass-weighted atomic displacements of atom $\alpha$ in Cartesian direction $i$,
\begin{equation}
    u_{\alpha i}(\bm{k}) = N^{-1/2}\sum_{\bm{m}} e^{-i\bm{k}\cdot\bm{R}_{\bm{m}\alpha}}\sqrt{m_\alpha}u_{\bm{m}\alpha i}\,,
\end{equation}
where $\bm{R}_{\bm{m}\alpha}$ is the equilibrium position of atom $\alpha$ in cell $\bm{m}$. 
The GF satisfies a Dyson equation,
\begin{equation}
    \bm{D}^{-1}(\bm{k},\omega) = \bm{D}_0^{-1}(\bm{k},\omega) - 2\bm{\Omega}(\bm{k})\bm{\pi}(\bm{k},\omega)\,,
\end{equation}
where $\bm{D}_0(\bm{k},\omega) = [\omega^2\bm{1} - \bm{\Omega}^2(\bm{k})]^{-1}$ is the harmonic GF of the lattice with phonon frequencies $\omega_\lambda(\bm{k})$, 
and $\bm{\pi}(\bm{k},\omega)$ is the self-energy. 
In our previous work~\cite{jasrasaria2024}, the phonon mode coefficients $c_{\alpha i,\lambda}(\bm{k})$ and frequencies $\omega_\lambda(\bm{k})$ were defined as eigenvectors and eigenvalues
of the dynamical matrix in the harmonic approximation,
\begin{equation}
    \bm{\mathcal{D}}_{\alpha i,\beta j}\left(\bm{k}\right)=\frac{1}{\sqrt{m_{\alpha}m_{\beta}}}\sum_{\bm{m}}e^{i\bm{k}\cdot\left(\bm{R}_{\bm{m}\alpha}-\bm{R}_{\bm{0}\beta}\right)}\frac{\partial^{2}\mathcal{V}}{\partial u_{\bm{m}\alpha i}\partial u_{\bm{0}\beta j}}\,,
\end{equation}
where derivatives of the lattice potential, $\mathcal{V}$, with respect to atomic displacements, $u_{\bm{m}\alpha i}=r_{\bm{m}\alpha i}-R_{\bm{m}\alpha i}$, are evaluated at the equilibrium lattice configuration. 
Here, we improve this harmonic description and use self-consistent phonon theory (SCP)~\cite{Hooton1958, KoehlerPRL1966, werthamer1970self, klein1972rise,tadano2018first} to obtain a more accurate temperature-dependent quasiparticle description of phonons. 
In our SCP calculations, the phonon modes and frequencies are defined via eigenvectors of
a modified dynamical matrix, $\bm{\mathcal{D}}(\bm{k}) + \bm{\mathcal{W}}(\bm{k})$, where $\bm{\mathcal{W}}(\bm{k})$ is the mean-field contribution due to quartic anharmonicity.



In VDMFT, the dynamics of the periodic lattice are mapped onto an ``impurity problem" that consists of a single unit cell coupled to a fictitious bath of harmonic oscillators through a tailored spectral density~\cite{GeorgesPhysRevB1992, GeorgesRevModPhys1996, kotliar2004strongly, KotliarRevModPhys2006, shih_anharmonic_2022, jasrasaria2024}. The local self-energy in a single unit cell, $\bm{\pi}(\omega)$, is calculated through the exact solution of the impurity problem and is used to approximate the self-energy of the lattice GF, $\pi_{\alpha i,\beta j}(\bm{k},\omega) \approx \pi_{\alpha i,\beta j}(\omega)$, thus providing a nonperturbative treatment of anharmonicity. The lattice GF and impurity problem are iteratively updated until self-consistency is achieved. We perform SCP calculations using the ALAMODE package~\cite{tadano2014anharmonic} (details given in the 
SM~\cite{Note1}) and use that quasiparticle basis in combination with VDMFT, following the approach described in detail in our previous work~\cite{shih_anharmonic_2022, jasrasaria2024}. As in our
previous work, the impurity GF is calculated by neglecting nuclear quantum effects and treating the nuclei classically.

Figure~\ref{fig1:model-vdmft}c illustrates the trace of the anharmonic spectral functions, $\bm{A}(\bm{k},\omega)=-\pi^{-1}\Im\bm{D}(\bm{k},\omega)$, of BaGG and SrGG calculated using SCP+VDMFT at 300\,K. Anharmonicity causes hardening of all phonon modes, especially those flat modes dominated by guest atoms rattling in the tetrakaidecahedral cages, which increase in frequency by 3\,cm$^{-1}$ and 12\,cm$^{-1}$ for BaGG and SrGG, respectively~\cite{jasrasaria2024}. These frequency shifts are accurately captured by the SCP quasiparticle basis, as shown by the black dashed lines in Fig.~\ref{fig1:model-vdmft}c. Finite lifetimes due to phonon-phonon scattering cause broadening of the quasiparticle dispersion, which is captured by the SCP+VDMFT spectral function. Again, the guest-dominated modes display significant broadening; in particular, the Sr rattling mode acquires a linewidth of 8\,cm$^{-1}$, which corresponds to a short lifetime of 0.67\,ps. The hardening and broadening of the rattling modes in these filled clathrates also affects their avoided crossing with the cage acoustic modes, thereby impacting the thermal conductivity of these materials.

We assess the accuracy of our SCP+VDMFT approach by comparing the spectral function of SrGG, the most anharmonic system studied here, with the exact spectral function computed using molecular dynamics (MD) simulations of a large supercell with periodic boundary conditions. The spectral functions at the $X$ point of the BZ are illustrated in Fig.~\ref{fig1:model-vdmft}d along with the frequency shifts and linewidths calculated on a 4$\times$4$\times$4 grid of the BZ, demonstrating excellent agreement between SCP+VDMFT and MD. We also compare linewidths to those calculated with lowest-order perturbation theory (PT) of three-phonon scattering processes using the SCP quasiparticle basis, which is the conventional approach for computing phonon linewidths~\cite{tadano_quartic_2018}. Details regarding the MD and SCP+PT calculations are given in the SM~\cite{Note1}. Unlike MD and SCP+VDMFT, Fig.~\ref{fig1:model-vdmft}d shows that SCP+PT inaccurately predicts negligible broadening for all modes in SrGG. This failure of SCP+PT has significant implications for calculations of the thermal conductivity, which rely on the accurate description of phonon linewidths, as described below.

\subsection{Thermal conductivities}

\subsubsection{Green-Kubo formalism}

The thermal conductivity, $\bm{\kappa}$, relates the macroscopic heat flux in a system, $\bm{J}$, with an applied temperature gradient. In the linear response regime, the thermal conductivity can be computed using the Kubo formula~\cite{kubo1957}, which depends on the autocorrelation function of the heat-flux operator:
\begin{subequations}
    \label{eq:kappaCorrelation}
\begin{align}
\kappa^{ij} &= \frac{V}{T} \lim_{\omega\rightarrow 0} 
    \frac{\mathrm{Im} \chi^{ij}(\omega+i\eta)}{\omega} \\
\chi^{ij}(t) &= \frac{i}{\hbar}\theta(t) 
    \langle \left[ J^i(t), J^j(0) \right] \rangle
\end{align}
\end{subequations}
where $V$ is the supercell volume. 
In the classical limit, this reduces to the familiar expression 
\begin{equation}
\label{eqn:kappaClassical}
    \kappa^{ij} = \frac{V}{k_BT^2} \int_0^\infty dt
        \langle J^i(t) J^j(0) \rangle,
\end{equation}
but here we aim to maintain quantum statistics, which is important for the low temperature conductivity.
In atomistic or lattice models, the heat flux operator is not uniquely defined. Here, we use the Hardy definition of the harmonic heat-flux operator~\cite{hardy_energy-flux_1963}, 
\begin{equation}
\begin{split}
    J^i &= \frac{\hbar}{2V} \sum_{\bm{k}}\sum_{\lambda,\lambda'} \omega_\lambda(\bm{k}) v^i_{\lambda,\lambda'}(\bm{k}) \\
    &\hspace{1em}\times \left[a^\dagger_\lambda(\bm{k})+a_\lambda(-\bm{k})\right]\left[a_{\lambda'}(\bm{k})-a^\dagger_{\lambda'}(-\bm{k})\right]\,,
\end{split}
\end{equation}
where $a^\dagger_{\lambda}(\bm{k})$ and $a_{\lambda}(\bm{k})$ are the creation and annihilation operators, respectively, of mode $\lambda$, and $\omega_{\lambda}(\bm{k})$ is its frequency. The generalized velocity matrix~\cite{caldarelli_many-body_2022} is
\begin{align}
    &\bm{v}_{\lambda,\lambda'}(\bm{k}) = \frac{1}{2\sqrt{\omega_\lambda(\bm{k})\omega_{\lambda'}(\bm{k})}} \nonumber \\
    &\quad\times \Bigg(\sum_{\alpha i,\beta j} c_{\alpha i,\lambda} (\bm{k}) \nabla_{\bm{k}} [\bm{\mathcal{D}}(\bm{k}) + \bm{\mathcal{W}}(\bm{k})]_{\alpha i,\beta j} c^*_{\beta j,\lambda'}(\bm{k})\Bigg)\,,
\end{align}
where, in the absence of degeneracies, the diagonal elements coincide with the usual phonon group velocities, $\bm{v}_{\lambda,\lambda} = \nabla_{\bm{k}}\omega_\lambda(\bm{k})$. 

While the response function, $\chi_{ij}(t)$, depends on products of four bosonic operators, or a two-particle GF, it can be approximated as a product of two single-particle GFs by neglecting vertex corrections~\cite{Mahan2000}. This approximation, which is exact in infinite dimensions~\cite{GeorgesPhysRevB1992}, is expected to be valid for materials with low thermal conductivity, in which resistive, Umklapp scattering processes are dominant over momentum-conserving, normal scattering processes that are characteristic of the hydrodynamic regime of thermal transport~\cite{guyer1966}. 
This approximation yields the thermal conductivity~\cite{Maradudin1962, PhysRevB.5.3909, isaeva_modeling_2019, caldarelli_many-body_2022},
\begin{align}
    &\kappa^{ij} = \frac{\pi\hbar}{VT}\sum_{\bm{k}}\sum_{\lambda,\lambda'}\omega_\lambda(\bm{k})\omega_{\lambda'}(\bm{k})v^i_{\lambda,\lambda'}(\bm{k})v^j_{\lambda',\lambda}(\bm{k}) \nonumber \\
    &\qquad\times \int_{-\infty}^\infty d\omega \left[\omega^2 + \omega^2_\lambda(\bm{k})\right]A_{\lambda,\lambda}(\bm{k},\omega)A_{\lambda',\lambda'}(\bm{k},\omega) \nonumber \\
    & \qquad\qquad\qquad \times \Bigg[-\frac{\partial n_T(\omega)}{\partial \omega}\Bigg]\,,
    \label{eq:kappaGK}
\end{align}
where 
$n_T(\omega) = \big(e^{\hbar\omega/k_BT} - 1\big)^{-1}$ is the Bose-Einstein distribution. 
%
Equation~(\ref{eq:kappaGK}) contains both intraband ($\lambda=\lambda'$) and interband ($\lambda\neq \lambda'$) contributions~\cite{simoncelli_unified_2019, isaeva_modeling_2019, simoncelli2022wigner, caldarelli_many-body_2022}.
The interband contributions become important in materials for which phonon broadening is on the same order as energy spacing between phonon bands, as seen in the clathrates studied here, especially SrGG (Fig.~\ref{fig1:model-vdmft}c).


\subsubsection{Boltzmann transport equation within the single mode approximation}

\begin{figure*}[ht!]
\includegraphics[width=5.5in]{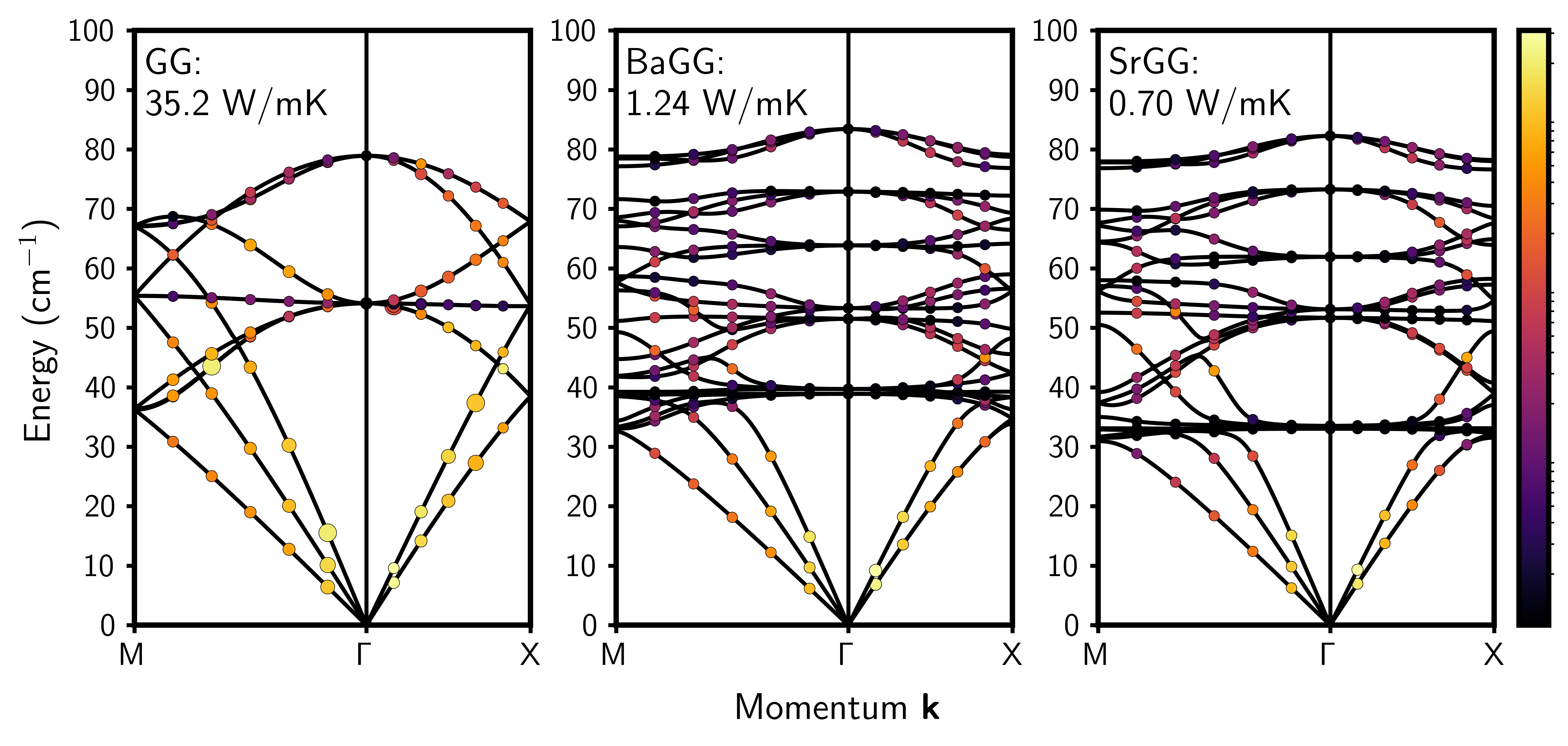}
\caption{Intraband contributions to the thermal conductivities of GG (left), BaGG (center), and SrGG (right) calculated at 300~K using the Green-Kubo formalism with spectral functions calculated using SCP+VDMFT. Points indicate the mode-resolved intraband thermal conductivity, where the size and color of each point correspond to the relative magnitude. Black lines indicate the SCP dispersion at 300~K. The top left of each panel gives the value of the total intraband thermal conductivity at 300~K.}
\label{fig3:kappaDispersion}
\end{figure*}

The full expression for the thermal conductivity given by Eq.~(\ref{eq:kappaGK}) can be simplified under the assumption that the phonon quasiparticle picture is valid such that interband transitions are negligible and that the phonon spectral densities have Lorenztian lineshapes centered at the harmonic phonon frequencies, $\omega_\lambda(\bm{k})$, with widths $\Gamma_\lambda(\bm{k})$. These approximations yield an expression for the thermal conductivity that
is identical to the BTE-SMA expression in Eq.~(\ref{eq:kappa_bte-sma}), 
with $v_\lambda^i(\bm{k}) = v_{\lambda,\lambda}^i(\bm{k})$
and $\tau_\lambda(\bm{k}) = 1/2\Gamma_\lambda(\bm{k})$.
%
%
%

The reliability of this BTE-SMA approach depends on the accuracy of the above approximations and of the frequencies and lifetimes used. For example, the frequencies and lifetimes can be calculated by a combination of mean-field and perturbation theory, as is commonly done, or extracted from our SCP+VDMFT approach, as shown in Fig.~\ref{fig1:model-vdmft}d.
Results obtained using both approaches will be compared in the following sections.

\section{Results and Discussion}

\subsection{Intraband transport}

First, we examine the intraband contributions to the thermal conductivity. We calculate the thermal conductivity at 300~K using the Green-Kubo formalism [Eq.~(\ref{eq:kappaGK})] with anharmonic spectral functions calculated with SCP+VDMFT sampled on a 12$\times$12$\times$12 Gamma-centered grid of the BZ and consider only the terms where $\lambda=\lambda'$. The intraband thermal conductivity of the empty clathrate, GG, is 35.2~W/mK, which is more than 20 times greater than that of the filled clathrates BaGG and SrGG, which have values of 1.24~W/mK and 0.70~W/mK, respectively. BaGG features weaker cage-guest interactions than SrGG, indicating the role of cage-guest anharmonicity in reducing the thermal conductivity.

Figure~\ref{fig3:kappaDispersion} illustrates the SCP dispersions of GG, BaGG, and SrGG at 300~K along a high-symmetry path through the BZ as well as each phonon mode's contribution to the intraband thermal conductivity. In GG, acoustic modes have the largest contributions to the thermal conductivity due to both their high phonon mode velocities and relatively long lifetimes. Optical modes contribute, as well, especially those near the zone edges with larger group velocities. In the filled clathrates, flat optical modes corresponding to rattling motions of X(2) guest atoms inside 24-atom cages, where X=Ba,Sr, cut through the acoustic modes of the cage lattice, leading to the avoided crossings that are characteristic of these materials~\cite{jasrasaria2024}. Consequently, the acoustic modes of the filled clathrates have diminished contributions to the thermal conductivity, especially as they approach the avoided crossing near the BZ edges. As seen in Fig.~\ref{fig3:kappaDispersion}, points corresponding to the flat, guest-dominant modes have negligible contributions to the thermal conductivity. 

The SM~\cite{Note1} shows the calculated mean-free-paths of acoustic modes of GG, BaGG, and SrGG at various temperatures. At all temperatures, the filled clathrates have mean-free-paths that are an order of magnitude or more smaller than those of the empty clathrate. This reduction is due to a sharp decrease in phonon lifetimes for the filled clathrate acoustic modes in a wide frequency range, confirming the role of nonresonant scattering in these materials~\cite{tadano_impact_2015}. In contrast, the acoustic phonon mode velocities are unchanged near the BZ center, and they only decrease as they approach the zone edge and gain significant character of the X(2) guest atoms.

\subsection{Interband transport}

\begin{figure*}[ht]
\includegraphics[width=5.06in]{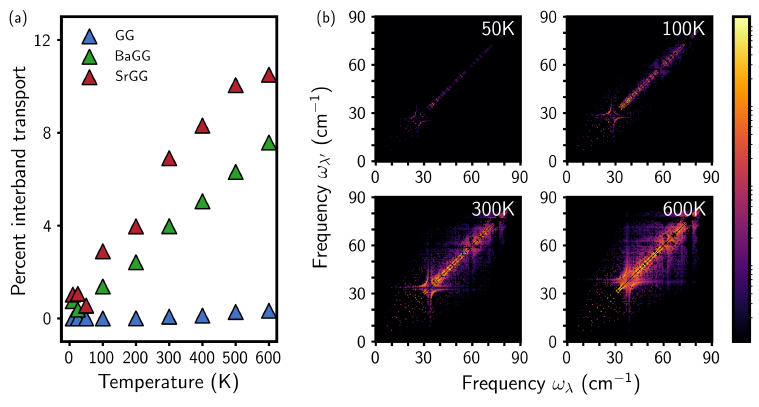}
\caption{(a) Interband contributions to the thermal conductivities of GG, BaGG, and SrGG calculated at different temeratures using the Green-Kubo formalism. (b) Interband thermal conductivities of SrGG showing transport between modes with frequencies $\omega_\lambda$ and $\omega_{\lambda\prime}$.}
\label{fig4:interband}
\end{figure*}

Due to the strong anharmonicity and significant mode mixing of the filled clathrates, especially near the avoided crossing of the cage-acoustic and guest-optical modes, we investigate the role of interband transport in the overall thermal conductivity~\cite{simoncelli_unified_2019, godse_anharmonic_2022}. Figure~\ref{fig4:interband}a shows the percent of the total thermal conductivity that is due to interband transport. As expected, GG has negligible interband transport at all temperatures because of its relatively harmonic lattice dynamics, for which the particle-like picture of thermal transport is dominant. The filled clathrates, however, have significant contributions from interband transport that increases with increasing temperature. At 600~K, 8\% and 11\% of the total thermal conductivity for BaGG and SrGG, respectively, are due to interband transport. We note that our coarse-grained model neglects flat, high-frequency modes related to intra-cage dynamics~\cite{tadano_impact_2015}, and contributions to interband transport may thus be more significant with an atomistic description of these clathrates~\cite{godse_anharmonic_2022}.

To better understand the mechanism underlying interband transport in SrGG, we analyze the mode-resolved interband contributions to the thermal conductivity at different temperatures, which are illustrated in Fig.~\ref{fig4:interband}b. There are few interband transitions at 50~K, and they increase slightly at 100~K, primarily occurring between phonon modes that are close in energy (i.e., $\omega_\lambda\sim\omega_{\lambda\prime}$). At 300~K, interband transport begins to occur more significantly between modes that are further apart in energy. In particular, a plus-shaped pattern appears around $\omega_\lambda = \omega_{\lambda\prime} \sim 32\,$cm$^{-1}$, which is the frequency of the flat rattling Sr(2) modes at 300K. These modes have broad linewidths of 8\,cm$^{-1}$ and and are intersected by several of the cage-acoustic modes (Fig.~\ref{fig1:model-vdmft}c). The overlapping spectral densities of the modes along these regions enable significant interband transport. Another plus-shaped region appears around 60\,cm$^{-1}$, which corresponds to the overlapping of the cage-optical and higher-energy X(2)-dominant modes at that energy. Further interband transport between modes with larger energy differences occurs at 600~K, and additional plus-shaped regions appear around the X(1) guest modes at 80\,cm$^{-1}$. As the X(2) rattling mode continues to harden and broaden at increased temperatures, the interband transitions around that frequency range become more prominent.

This analysis directly demonstrates the role of guest atoms in enabling interband transport, which is negligible in the empty clathrates. However, we emphasize that while a significant percentage of the total thermal conductivity is due to interband transport enabled by guest-dominant modes, the absolute value of the interband thermal conductivity is small (the interband thermal conductivity of BaGG and SrGG at 300~K is 0.051~W/mK and 0.053~W/mK, respectively), and the primary consequence of the guest atoms are to reduce the overall thermal conductivity. Despite the differences in anharmonicity between BaGG and SrGG, which is reflected by differences in their spectral functions (Fig.~\ref{fig1:model-vdmft}c), they have similar values for interband transport at 300~K. While the broad SrGG rattling modes have more spectral overlap with acoustic modes than those of BaGG, which enables interband transfer, they also have shorter lifetimes and lower frequencies, which reduce those transitions' contributions to the thermal conductivity. These results indicate that a complex balance of factors and spectral features underlie the thermal conductivity, which is characterized by a single value; although brute force molecular dynamics is commonly applied to calculate the thermal conductivity as the integral of the heat flux autocorrelation function [Eq.~(\ref{eqn:kappaClassical})], such an approach lacks the atomistic insights available from the mode-resolved Green's function approach taken here.

\subsection{The role of nonperturbative effects on thermal conductivity}

\begin{figure*}[ht]
\includegraphics[width=5.06in]{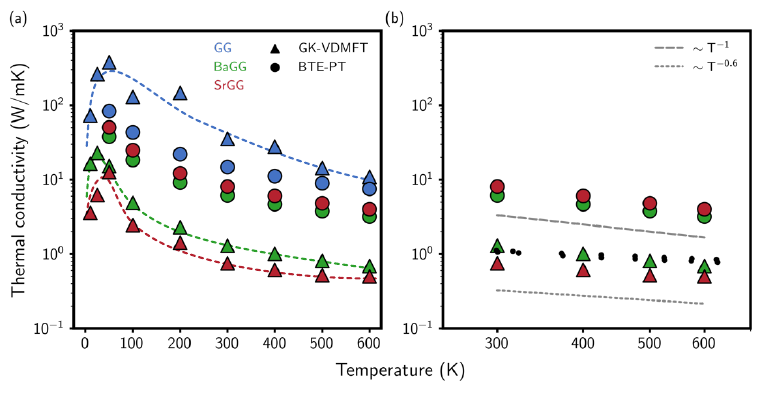}
\caption{(a) Thermal conductivities of GG (blue), BaGG (green), and SrGG (red) calculated using the Green-Kubo formalism [Eq.~(\ref{eq:kappaGK})] with spectral functions calculated using SCP+VDMFT (triangles) and using the BTE formalism [Eq.~(\ref{eq:kappa_bte-sma})] with lifetimes calculated using SCP+PT (circles). All calculations are sampled on a $12\times 12 \times 12$ Gamma-centered grid of the BZ. Dashed lines are a guide to the eye. (b) Thermal conductivities at higher temperatures, where black points are experimental measurements from Ref.~\citep{may_characterization_2009}. Grey lines indicating different power laws are a guide to the eye.}
\label{fig2:temperature}
\end{figure*}

To better understand the role of nonperturbative effects on the thermal conductivity, we compare the thermal conductivity across materials and temperatures computed using two different methods: the Green-Kubo formalism [Eq.~(\ref{eq:kappaGK})] with spectral functions calculated using SCP+VDMFT (GK-VDMFT) and the BTE-SMA approach [Eq.~(\ref{eq:kappa_bte-sma})] with SCP phonons and lifetimes calculated using conventional lowest-order PT of the three-phonon scattering processes (BTE-PT). Details regarding the BTE-PT calculations are provided in the SM~\cite{Note1}.

When calculated using GK-VDMFT, the thermal conductivity for all materials peaks around 25-50~K and then decreases with increasing temperature. As discussed above, the introduction of guest atoms decreases the thermal conductivity by more than one order of magnitude, and BaGG, which features weaker cage-guest anharmonicity, has a larger thermal conductivity than SrGG at all temperatures. The BTE-PT approach results in thermal conductivity values that are consistent with GK-VDMFT for the quasi-harmonic GG, but they disagree significantly for the filled clathrates. BTE-PT predicts only a factor of two reduction between GG and BaGG, and it incorrectly predicts that the more anharmonic SrGG has a larger thermal conductivity than BaGG. These discrepancies are a result of the failure of SCP+PT to capture the short lifetimes of the filled clathrates (Fig.~\ref{fig1:model-vdmft}d), causing a significant overestimation of their thermal conductivities. As shown in the SM~\cite{Note1}, using the BTE-SMA formalism with lifetimes determined from SCP+VDMFT calculations yields values that closely align with those calculated by the GK-VDMFT approach, suggesting that the success of the BTE-SMA depends strongly on the quality of the phonon lifetimes used and, in this case, is not due to the exclusion of interband transport. The SM~\cite{Note1} also illustrates comparisons to thermal conductivities calculated in the classical limit [Eq.~(\ref{eqn:kappaClassical})] using molecular dynamics simulations, showing consistency with the GK-VDMFT results. These comparisons suggest that the ultralow thermal conductivities of the filled clathrates result from four-phonon and higher-order scattering processes that are not described by SCP+PT but that are inherently included in our SCP+VDMFT approach, highlighting the importance of nonperturbative anharmonic effects on observables like the thermal conductivity. In Fig.~\ref{fig2:temperature}b, we compare our calculated values of the thermal conductivity for BaGG to experimental data, focusing on high-temperature data, where intrinsic effects of anharmonicity dominate over effects due to grain boundary scattering, dynamical disorder, and nuclear quantum effects. We see excellent agreement although the degree of quantitative agreement needs to be further understood in light of the approximations made in our coarse-grained model. At lower temperatures, phonon-grain boundary scattering can become important~\cite{tadano_impact_2015}, as discussed in the SM~\cite{Note1}.

Finally, we consider the temperature dependence of the total thermal conductivity (Fig.~\ref{fig2:temperature}b), as strong interactions can produce anomalous  behavior, as in electronic transport properties~\cite{gunnarsson_colloquium_2003}. Here, we focus on the high-temperature dependence, as our calculations neglect nuclear quantum effects and grain boundary scattering, which may affect the behavior of thermal conductivities at low temperatures. For all materials studied here, BTE-PT predicts a $T^{-1}$ temperature dependence of the thermal conductivity. This temperature dependence is dictated by that of the phonon lifetimes, for which a $T^{-1}$ scaling is inherent to the PT formalism at temperatures greater than the phonon frequency (see SM~\cite{Note1}). In contrast, the GK-VDMFT approach predicts different high-temperature behavior for the different clathrates. BaGG has an intermediate scaling of $\sim T^{-0.9}$, and SrGG, the most anharmonic system, has the most moderate scaling, $\sim T^{-0.6}$. While interband transport can account for deviations from the $T^{-1}$ scaling in some materials~\cite{simoncelli2022wigner}, the good agreement between the GK-VDMFT and BTE-VDMFT approaches (illustrated in the SM~\cite{Note1}) suggest that interband transport is not crucial to this behavior in the clathrate models studied here.

\begin{figure}[ht!]
\includegraphics[width=3.375in]{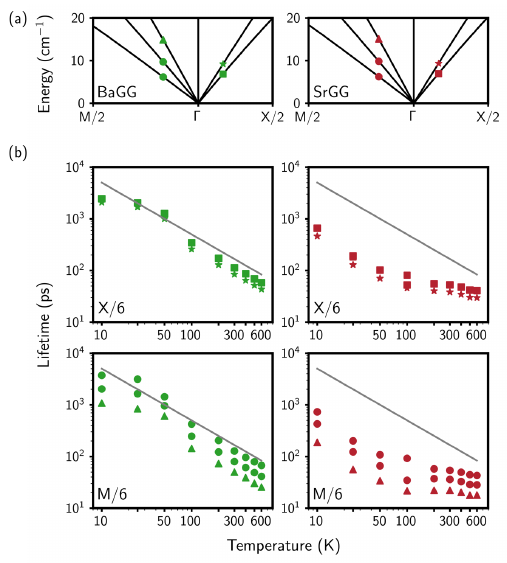}
\caption{(a) Renormalized dispersion of acoustic modes near the BZ center of BaGG (left) and SrGG (right) calculated using SCP at 300\,K. (b) The acoustic mode lifetimes of BaGG (left) and SrGG (right) extracted from the self-energy calculated using the SCP-VDMFT approach. Grey lines indicating a $T^{-1}$ power law are a guide to the eye.}
\label{fig5:modeLifetimes}
\end{figure}

The high-temperature dependence of the thermal conductivity primarily arises from that of the acoustic phonon lifetimes. Thus, we analyze the temperature dependence of longitudinal and transverse acoustic phonon modes near the BZ center of BaGG and SrGG, as illustrated in Fig.~\ref{fig5:modeLifetimes}. While the mode lifetimes of BaGG largely follow a $T^{-1}$ behavior, the mode lifetimes of SrGG deviate significantly from the $T^{-1}$ behavior that is predicted by perturbation theory and begin to saturate to a milder scaling at 50~K. This high-temperature saturation of the phonon lifetimes, and thus thermal conductivities, is reminiscent of ``resistivity saturation" in metals~\cite{gunnarsson_colloquium_2003, millis_resistivity_1999, jaramillo_origins_2014, fetherolf_conductivity_2023} and is a hallmark of nonperturbative interactions. This result indicates that the acoustic modes of the SrGG cage framework are subject to multiphonon scattering processes that are not captured by conventional PT. Despite their relatively narrow linewidths and long lifetimes (Fig.~\ref{fig1:model-vdmft}), these acoustic phonons' low energies and strong coupling to Sr(2) rattling modes makes them more likely to participate in multiphonon scattering processes. Thus, we find that including nonperturbative descriptions of anharmonicity is essential in accurately capturing thermal transport properties, especially at higher temperatures where multiphonon processes are activated.

\section{Conclusions}

In conclusion, we have developed a framework based on VDMFT for the efficient and accurate calculation of thermal transport properties in real materials, including nonperturbative anharmonic effects and both intraband and interband transport mechanisms. We have applied this methodology to understand the temperature-dependent thermal conductivities of type-I clathrate solids. Although we have used a coarse-grained representation of the atomic structure of these type-I clathrates, our results for the thermal conductivity are qualitatively and quantitatively consistent with \textit{ab initio} calculations. The authors of Ref.~\onlinecite{tadano_quartic_2018} calculate a value of 0.97~W/mK for BaGG at 300~K using the BTE-SMA approach with lifetimes extracted from SCP+PT calculations and ignoring interband transport, and the authors of Ref.~\onlinecite{godse_anharmonic_2022} calculate values of 1.26~W/mK and 0.77~W/mK for BaGG and SrGG, respectively, using a similar approach while also including interband transport. Those works successfully capture the reduction in thermal conductivities of the filled clathrates, even while using conventional PT, because of the atomistic detail in their clathrate structures (Fig.~\ref{fig1:model-vdmft}a), which enables three-phonon processes that are strictly zero in our coarse-grained structure (Fig.~\ref{fig1:model-vdmft}b).
Resolving this discrepancy, wherein quite different approaches predict similar conductivities, is an important goal of our future research, which will be enabled by our fully atomistic, first-principles implementation of VDMFT.
However, results with our coarse-grained model do confirm that the mechanism for thermal conductivity reduction in the filled clathrates is due to nonresonant scattering that decreases lifetimes of cage acoustic modes~\cite{tadano_impact_2015, tadano_quartic_2018, godse_anharmonic_2022}, even at frequencies away from those of the guest rattling modes, and that changes in acoustic mode velocity play a relatively minor role.

Finally, as shown in Fig.~\ref{fig2:temperature}, our calculations are able to capture the experimental high-temperature behavior of BaGG, while the results of Refs.~\onlinecite{tadano_impact_2015, tadano_quartic_2018} show a steeper scaling with increasing temperature. This discrepancy may be due to the role of nonperturbative effects discussed above. 

Nonperturbative effects, in the form of a phonon Kondo effect, have also been suggested as being responsible for some unconventional behaviors of the thermal conductivity of type-I clathrates~\cite{ikeda_kondo-like_2019}, including a narrow maximum at temperatures lower than the Debye temperature and a larger high-temperature conductivity than predicted by perturbation theory. 
Given the success of DMFT and its associated impurity problem in explaining
Kondo physics in electronic systems, our method is uniquely suited to capture any such physics. 
The hybridization of a dispersionless and dispersive band is indeed suggestive, although the exact expectations of a phonon Kondo effect are unclear, given that phonons lack a Fermi energy and exhibit a trivial maximum in their conductivity (see SM~\cite{Note1} for additional details).
Although the Debye temperature of our studied clathrates is 60--100~K, our calculated conductivities exhibit a maximum at lower temperatures of 25--50~K, which are comparable to the Einstein temperature of the rattling modes. 
Moreover, at high temperature we see a weaker dependence of the conductivity than predicted by perturbation theory. Both of these results are qualitatively in line with experimental observations.


The results of this work can be used to extract design principles for the development of real clathrate materials. The introduction of guest atoms inside clathrate cages increases anharmonicity of the lattice dynamics and lowers the thermal conductivity significantly. Smaller guest atoms, such as Sr compared to Ba, have more free space to move within the cages, resulting in more anharmonic cage-guest interactions as well as lower-frequency X(2) rattling modes with shorter lifetimes due to multiphonon scattering. These factors generally reduce the thermal conductivity, as shown in Fig.~\ref{fig2:temperature}a. However, these factors can also enable interband transport, especially at higher temperatures, which can elevate the thermal conductivity (Fig.~\ref{fig4:interband}a). This competition indicates that a careful trade-off must be considered when designing thermoelectric materials with low thermal conductivities.

In future work, we will apply our VDMFT framework to a fully atomistic description of the type-I clathrate materials to evaluate the relative importance of three-phonon versus multi-phonon scattering processes in real materials with lower symmetry than our coarse-grained model.
Moreover, although our calculated conductivity partially captures nuclear quantum effects through its use of Bose-Einstein statistics, future work will study the role of nuclear quantum dynamics, which may be important at very low temperatures.

\textit{Acknowledgements.}
This work
was supported by the U.S. Department of Energy, Office
of Science, Basic Energy Sciences, under Award No. DESC0023002. We acknowledge computing resources from
Columbia University’s Shared Research Computing Facility project, which is supported by NIH Research Facility
Improvement Grant No. 1G20RR030893-01, and associated
funds from the New York State Empire State Development,
Division of Science Technology and Innovation (NYSTAR),
Contract No. C090171
The Flatiron Institute is a division of the Simons Foundation.

%

\end{document}


\title{Supplemental Material: Strong anharmonicity dictates ultralow thermal conductivities of type-I clathrates}

\author{Dipti Jasrasaria}
\email{dj2667@columbia.edu}
\affiliation{Department of Chemistry, Columbia University, New York, New York 10027, USA}

\author{Timothy C. Berkelbach}
\email{t.berkelbach@columbia.edu}
\affiliation{Department of Chemistry, Columbia University, New York, New York 10027, USA}
\affiliation{Initiative for Computational Catalysis, Flatiron Institute, New York, New York 10010, USA}

\date{\today}

\maketitle

\section{Clathrate model and parameters}

Our clathrate model consists of large ``cage" atoms that form a face-centered cubic (FCC) lattice with a lattice constant of $a=10.95$\,\AA~and with four sites in each unit cell, whose coordinates are collected in Table~S1. The cage atoms have a mass of $m_\text{cage}=1647.924$\,amu and interact with a Lennard-Jones (LJ) potential, which is defined by the parameters $\epsilon=2.7876$\,eV and $\sigma=7.0345$\,\AA~and a cutoff of $r_c=14.448$\,\AA.

\begin{center}
\begin{table}[h!]
    \caption{Fractional coordinates for the four sites in the FCC lattice unit cell. The lattice constant is $a=10.95$\,\AA.}
    \centering
    \begin{tabular}{cccc}
    \hline
       atom  & $x$ & $y$ & $z$ \\
    \hline
        $0$ & $0$ & $0$ & $0$ \\
        $1$ & $a/2$ & $a/2$ & $0$ \\
        $2$ & $a/2$ & $0$ & $a/2$ \\
        $3$ & $0$ & $a/2$ & $a/2$ \\
    \hline
    \end{tabular}
    \label{tab:fcc}
\end{table}
\par\end{center}

Additionally, at each lattice site is a smaller ``guest" atom that interacts with the cage atom at that site through an anharmonic, quartic potential. The parameters defining the quartic potential for the cage-guest interactions at each of the four FCC sites are given in Table~S2. The Ba guest atoms have mass $m_\text{Ba}=137.327$\,amu, and the Sr guest atoms have mass $m_\text{Sr}=87.62$\,amu. 

\begin{center}
\begin{table}[h!]
    \caption{Parameters defining the quartic cage-guest potential given by Eq.~(3) of the main text. The units for $K$ are eV$\cdot$\AA$^{-2}$, and the units for $g$ are eV$\cdot$\AA$^{-4}$}
    \centering
    \begin{tabular}{ccccccc}
    \hline
        atom & $K_x$ & $K_y$ & $K_z$ & $g_x$ & $g_y$ & $g_z$ \\
    \hline
        Ba$_0$ & 2.4820 & 2.4820 & 2.4820 & 0.6795 & 0.6795 & 0.6795 \\
        Ba$_1$ & 1.6563 & 0.6763 & 0.6763 & 0.8198 & 0.3550 & 0.3550 \\
        Ba$_2$ & 0.6763 & 0.6763 & 1.6563 & 0.3550 & 0.3550 & 0.8198 \\
        Ba$_3$ & 0.6763 & 1.6563 & 0.6763 & 0.3550 & 0.8198 & 0.3550 \\
    \hline
        Sr$_0$ & 1.6708 & 1.6708 & 1.6708 & 0.2293 & 0.2293 & 0.2293 \\
        Sr$_1$ & 0.9873 & 0.1483 & 0.1483 & 0.4555 & 0.2438 & 0.2438 \\
        Sr$_2$ & 0.1483 & 0.1483 & 0.9873 & 0.2438 & 0.2438 & 0.4555 \\
        Sr$_3$ & 0.1483 & 0.9873 & 0.1483 & 0.2438 & 0.4555 & 0.2438 \\
    \hline
    \end{tabular}
    \label{tab:Vq}
\end{table}
\par\end{center}

\section{Calculating phonon Green's functions with molecular dynamics}

The exact anharmonic dynamics of the lattice, in the classical limit assumed here, can be modeled using molecular dynamics (MD) simulations of a large supercell with periodic boundary conditions. We compute the anharmonic lattice GF:
\begin{equation}
\bm{D}_{\lambda,\lambda'}\left(\bm{k},t\right) = \frac{\theta(t)}{k_B T}\left\langle \dot{u}_{\lambda}\left(\bm{k},t\right)u_{\lambda'}\left(-\bm{k},0\right)\right\rangle\,.
\end{equation}
The GFs were calculated from simulations of supercells of $8\times 8\times 8$ unit cells with periodic boundary conditions by averaging over 20,000 trajectories of 20 ps each with a timestep of 0.0025 ps. Initial configurations were sampled at intervals of 25 ps from a MD trajectory, where the temperature was controlled using a Langevin thermostat. MD simulations were performed using the LAMMPS code~\cite{plimpton1995fast}. We fit each element of the GF using two parameters to the functional form given by
\begin{equation}
    \bm{D}_{\lambda,\lambda'}(\bm{k},t) = -\frac{\theta(t)}{\Omega_{\lambda,\lambda'}(\bm{k})} \exp\big(-\gamma_{\lambda,\lambda'}(\bm{k})t/2\big)\sin\big(\Omega_{\lambda,\lambda'}(\bm{k})t\big)\,,
\end{equation}
where $\Omega_{\lambda,\lambda'}(\bm{k}) = \omega^2_{\lambda,\lambda'}(\bm{k}) - \frac{1}{4}\gamma^2_{\lambda,\lambda'}(\bm{k})$. Then, we  analytically perform a Fourier transform to obtain the frequency-domain GF and then compute the spectral function.

\section{Self-consistent phonon calculations}

To calculate temperature-dependent frequencies within self-consistent phonon (SCP) theory~\cite{Hooton1958, KoehlerPRL1966, werthamer1970self, klein1972rise, tadano2018first}, we self-consistently solve the equation
\begin{equation}
    [\bm{\mathcal{D}}(\bm{k}) + \bm{\mathcal{W}}(\bm{k})]_{\lambda,\lambda'} = \omega^2_\lambda(\bm{k}) \delta_{\lambda,\lambda'} + \frac{\hbar}{2} \sum_{\bm{k}'}\sum_{\lambda ''} \Phi(\bm{k}\lambda,-\bm{k}\lambda ',\bm{k}'\lambda'',-\bm{k}'\lambda'') \frac{1+2n\big(\Omega_{\lambda''}(\bm{k}')\big)}{2\Omega_{\lambda''}(\bm{k}')}\,.
\end{equation}
Here, $\omega_\lambda(\bm{k})$ is the harmonic frequency of mode $\lambda$ that comes from diagonalizing the dynamical matrix, $\bm{\mathcal{D}}(\bm{k})$, and $\Omega_\lambda(\bm{k})$ is the temperature-dependent renormalized frequency of mode $\lambda$ that comes from diagonalizing $[\bm{\mathcal{D}}(\bm{k})+\bm{\mathcal{W}}(\bm{k})]$. $n(\omega) = [\exp(\hbar\omega/k_BT)-1]^{-1}$ is the Bose-Einstein distribution, and $\Phi(\bm{k}\lambda,-\bm{k}\lambda ',\bm{k}'\lambda'',-\bm{k}'\lambda'')$ is the reciprocal representation of the fourth-order interatomic force constant (IFCs) computed using the harmonic eigenvectors of the dynamical matrix.

To obtain the IFCs for each clathrate system, we first fit the harmonic IFCs using the finite displacement approach~\cite{Parlinski1997}. To generate the reference data set, we displace atoms from their equilibrium positions in a 3$\times$3$\times$3 supercell by lengths of 0.01\,\AA, 0.02\,\AA, and 0.03\,\AA. Then, we simultaneously estimate anharmonic IFCs up to sixth order using the compressive sensing lattice dynamics method~\cite{Zhou2014}, using a reference data set of 1000 configurations sampled from an equilibrium MD trajectory of a 3$\times$3$\times$3 supercell at 600~K. The temperature was controlled using a Langevin thermostat, and configurations were sampled at intervals of 2.5\,ps to ensure that they were not correlated. We consider all harmonic and anharmonic IFCs between cage atoms within the cutoff radius of the LJ potential ($r_c=14.448$\,\AA) and cage-guest pairs at the same lattice site.

Following Ref.~\onlinecite{tadano_quartic_2018}, we perform SCP calculations at $\bm{k}'=\bm{0}$, and we use Fourier interpolation to obtain the anharmonic frequencies and eigenvectors at arbitrary points in the Brillouin zone. All IFC estimation and SCP calculations were performed using the ALAMODE package~\cite{tadano2014anharmonic}.

\section{Calculating phonon linewidths using perturbation theory\label{sec:linewidths_PT}}

To calculate phonon lifetimes using lowest-order perturbation theory (PT) of three-phonon scattering processes using the SCP quasiparticle basis~\cite{tadano_quartic_2018}, we calculate
\begin{align}
    \Gamma_\lambda(\bm{k},\omega) = &\frac{\pi}{2N} \sum_{\bm{k}',\bm{k}''} \sum_{\lambda',\lambda''} \big\vert V_{\lambda,\lambda',\lambda''}^{(3)} (-\bm{k},\bm{k}',\bm{k}'')\big\vert^2 \nonumber \\
    &\times\bigg[ \bigg(n\big(\Omega_{\lambda'}(\bm{k}')\big) + n\big(\Omega_{\lambda''}(\bm{k}'')\big) + 1\bigg)\delta\bigg(\omega-\Omega_{\lambda'}(\bm{k}')-\Omega_{\lambda''}(\bm{k}'')\bigg) \nonumber \\
    &-2\bigg(n\big(\Omega_{\lambda'}(\bm{k}')\big) - n\big(\Omega_{\lambda''}(\bm{k}'')\big)\bigg)\delta\bigg(\omega-\Omega_{\lambda'}(\bm{k}')+\Omega_{\lambda''}(\bm{k}'')\bigg)\bigg]\,,
    \label{eqn:linewidthPT}
\end{align}
where $\Omega_\lambda(\bm{k})$ is the anharmonic frequency of mode $\lambda$ calculated using SCP. The three-phonon scattering matrix element is given by
\begin{equation}
    V_{\lambda,\lambda',\lambda''}^{(3)}(\bm{k},\bm{k}',\bm{k}'') = \bigg(\frac{\hbar^3}{8N^2\Omega_\lambda(\bm{k})\Omega_{\lambda'}(\bm{k}')\Omega_{\lambda''}(\bm{k}'')}\bigg)^{1/2} \Phi(\bm{k}\lambda, \bm{k}'\lambda', \bm{k}''\lambda'')\,,
\end{equation}
where $\Phi(\bm{k}\lambda, \bm{k}'\lambda', \bm{k}''\lambda'')$ is the reciprocal representation of the third-order IFC computed using the anharmonic eigenvectors.

Linewidths illustrated in Fig.~1 of the main text were calculated using a 16$\times$16$\times$16 $\Gamma$-centered $\bm{k}$-point grid and the Dirac delta functions were evaluated using the tetrahedron method. Again, all PT calculations were performed using the ALAMODE package~\cite{tadano2014anharmonic}.

From the linewidths, the phonon lifetimes can be computed as $\tau_\lambda(\bm{k}) = 1/2\Gamma_\lambda(\bm{k})$. The temperature-dependence of the linewidths enters primarily through the Bose-Einstein distribution [Eq. (\ref{eqn:linewidthPT})], which can be approximated as $n(\omega)\sim\hbar\omega/k_BT$ when $k_BT\gg\hbar\omega$. Thus, perturbation theory dictates that lifetimes have a high-temperature dependence of $\tau\sim1/T$.  

\section{Phonon mean-free-paths}

\begin{figure*}[ht]
\includegraphics[width=6.75in]{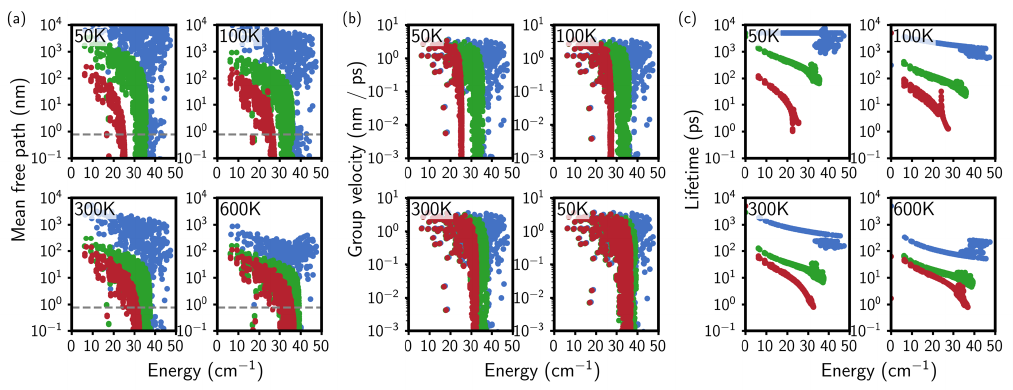}
\caption{(a) Phonon mean-free-paths of acoustic modes of GG (blue), BaGG (green), and SrGG (red) calculated as $\ell_\lambda(\bm{k})=v_\lambda(\bm{k})\tau_\lambda(\bm{k})$, where $v_\lambda(\bm{k})$ is the group velocity, and $\tau_\lambda(\bm{k})$ is the lifetime. The grey dashed line indicates the Ioffe-Regel limit, given by the cage-cage nearest neighbor distance. (b) Group velocities of acoustic modes of GG (blue), BaGG (green), and SrGG (red). (c) Lifetimes of acoustic modes of GG (blue), BaGG (green), and SrGG (red) calculated from the SCP+VDMFT self-energy, as described in Fig.~1d of the main text.
}
\label{figS1:mfp}
\end{figure*}

\section{Calculating thermal conductivities using molecular dynamics simulations}

The anharmonic dynamics of the lattice can be modeled using MD simulations of a large supercell with periodic boundary conditions. From these simulations, the thermal conductivity can be computed directly using the classical limit of Eq.~(11) of the main text, which is given by~\cite{kubo1957}
\begin{equation}
    \kappa^{ij}=\frac{V}{k_BT^2}\int_0^\infty dt \langle J^i(t)J^j(0)\rangle\,,
    \label{eqn:kappaClassical}
\end{equation}
where the Hardy harmonic heat flux~\cite{hardy_energy-flux_1963} is given by
\begin{equation}
    \bm{J}(t)=-\frac{1}{2V}\sum_{\bm{m}\alpha,\bm{n}\beta}\bigg(\sum_{ij}\frac{\partial^2\bm{\mathcal{V}}}{\partial u_{\bm{m}\alpha i} \partial u_{\bm{n}\beta j}} v_{\bm{m}\alpha i}(t) u_{\bm{n}\beta j}(t)\bigg) (\bm{R}_{\bm{m}\alpha} - \bm{R}_{\bm{n}\beta})\,,
\end{equation}
where $v_{\bm{m}\alpha i}(t)$ and $u_{\bm{m}\alpha i}(t)$ are the $i$ component of the velocity and displacement, respectively, of atom $\alpha$ in cell $\bm{m}$, and $\bm{R}_{\bm{m}\alpha}$ is the equilibrium position of atom $\alpha$ in cell $\bm{m}$. This expression ignores conductive contributions to the heat flux, which are negligible in solids.

The heat-flux autocorrelation functions were calculated from simulations of supercells of $4\times 4\times 4$ unit cells with periodic boundary conditions, which has shown to be large enough for convergence~\cite{tretiakov_thermal_2004}, by averaging over 10,000 trajectories of 1~ns each with a timestep of 0.0025~ps. Initial configurations were sampled at intervals of 25~ps from a MD trajectory, where the temperature was controlled using a Langevin thermostat. All MD simulations were performed using the LAMMPS code~\cite{plimpton1995fast}.

\begin{figure*}[ht]
\includegraphics[width=6.75in]{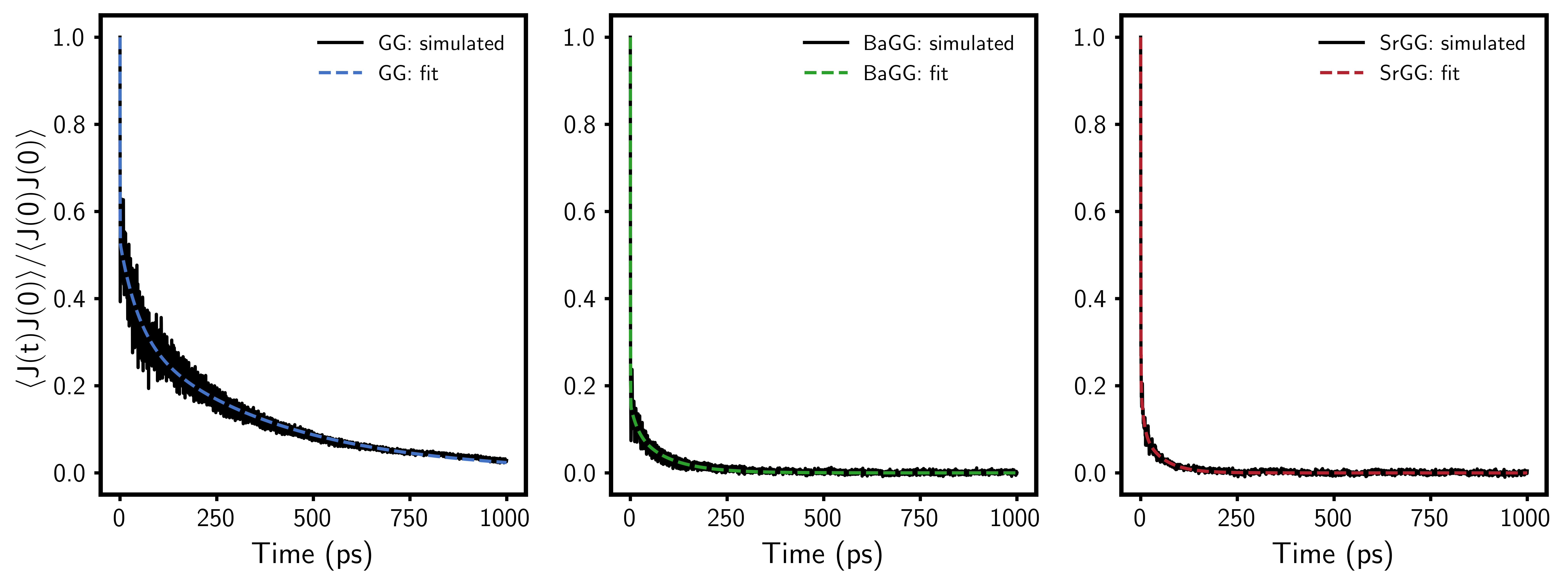}
\caption{The heat-flux autocorrelation function simulated using MD simulations at 300\,K for GG (left), BaGG (center), and SrGG (right) as well as fits according to Eq. (\ref{eqn:mdFit}).
}
\label{figS2:mdFit}
\end{figure*}

To facilitate convergence of the integral, which requires the propagation of several long trajectories as the autocorrelation function decays quite slowly, we fit the simulated correlation function to a triexponential functional form~\cite{dong_theoretical_2001, tretiakov_thermal_2004}:
\begin{equation}
    \frac{\langle J(t)J(0) \rangle}{\langle J(0) J(0) \rangle} = A_1 \exp(-t/\tau_1) + A_2\exp(-t/\tau_2) + A_3\exp(-t/\tau_3)\,,
    \label{eqn:mdFit}
\end{equation}
where the coefficients $A_1,A_2,A_3$ are constrained to sum to 1, and perform the Fourier transform analytically. Examples of heat-flux autocorrelation functions and their fits are given in Fig.~\ref{figS2:mdFit}. Note that, even with this fitting technique, the thermal conductivities calculated directly using MD simulations have large error bars due to statistical fluctuations that affect, for example, the initial value of the heat-flux autocorrelation function.

\begin{figure}[ht]
\includegraphics[width=3.38in]{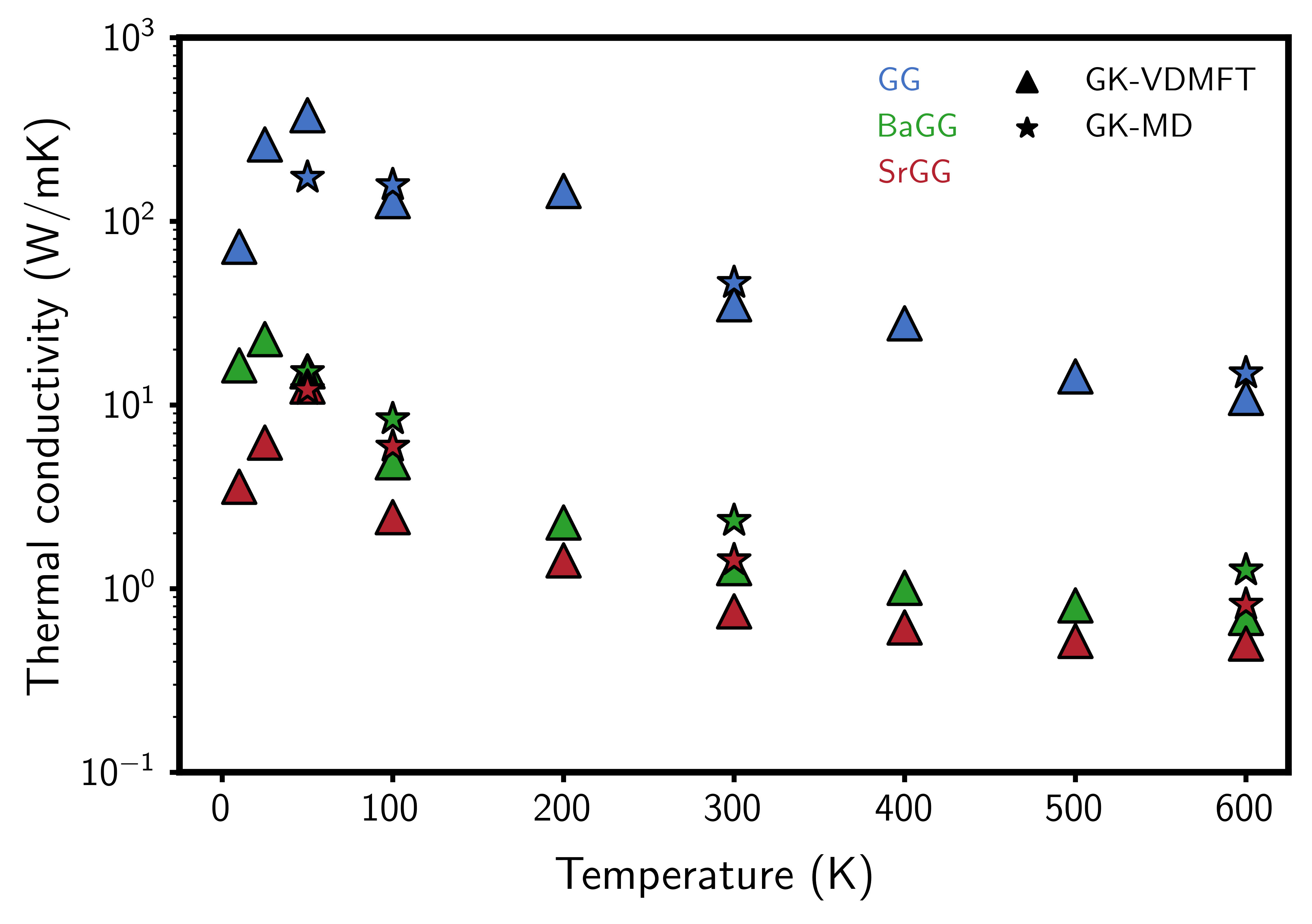}
\caption{Thermal conductivities of GG (blue), BaGG (green), and SrGG (red) calculated using the Green-Kubo formalism [Eq. (14) of the main text] with
spectral functions calculated using SCP+VDMFT (triangles) and using the classical Green-Kubo formalism [Eq. (\ref{eqn:kappaClassical})] with the heat-flux calculated using MD simulations
(stars).}
\label{figS3:kappaMD}
\end{figure}

The thermal conductivities calculating using MD simulations for all three clathrate systems at various temperatures are illustrated in Fig.~\ref{figS3:kappaMD}. The thermal conductivities calculated from MD simluations are in good agreement with those calculated using the GK-VDMFT approach. These calculations of the thermal conductivity, while classical, do include the description of normal-conserving scattering processes that are neglected in both the Green Kubo and BTE approaches given by Eqs.~(14) and (1), respectively, of the main text, validating the approximation to neglect vertex corrections.

\newpage

\section{Calculating thermal conductivities using the Boltzmann Transport Equation within the single mode approximation}

The thermal conductivity can be calculated using an expression based on the Boltzmann transport equation within the
single-mode approximation (BTE-SMA), which is given by Eq.~(1) of the main text. The mode frequencies and group velocities are calculated using the temperature-dependent SCP phonons.

The lifetimes used in the BTE-PT results were calculated as described in Sec.~\ref{sec:linewidths_PT} using a $12\times 12\times 12$ $\Gamma$-centered grid of the Brillouin zone. 

Additionally, the BTE-SMA formalism can be calculated using lifetimes determined from the SCP+VDMFT self-energy as described in Fig.~1d of the main text, which we call the BTE-VDMFT approach.

\begin{figure}[ht]
\includegraphics[width=3.38in]{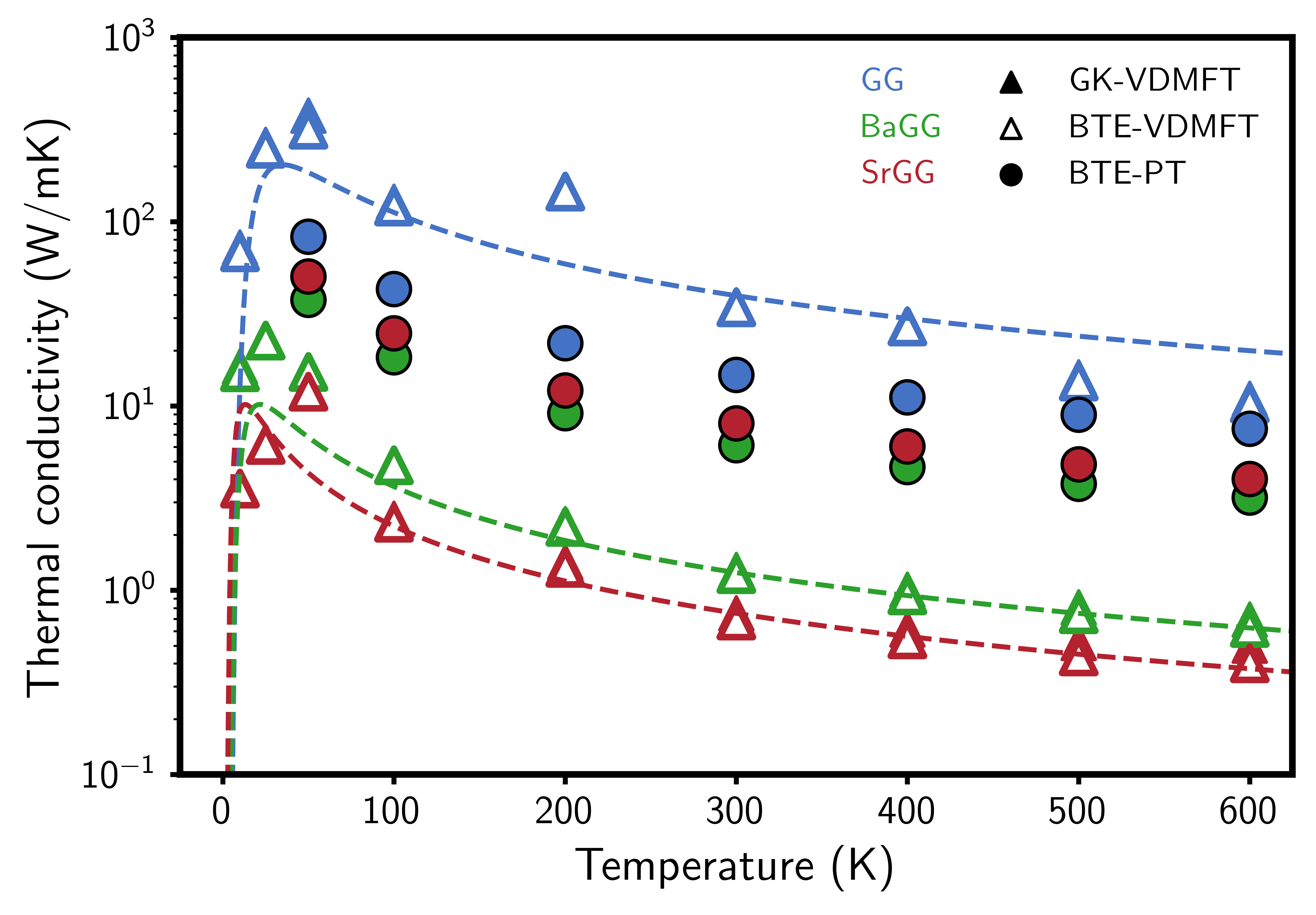}
\caption{Thermal conductivities of GG (blue), BaGG (green), and SrGG (red) calculated using the Green-Kubo formalism [Eq. (14) of the main text] with
spectral functions calculated using SCP+VDMFT (triangles) and using the BTE formalism [Eq. (1) of the main text] with lifetimes calculated from the SCP+VDMFT self-energy (open triangles) and using SCP+PT (circles). \textcolor{black}{Note some of the closed triangles are hidden by the open triangles, indicating the close agreement between the GK-VDMFT and BTE-VDMFT approaches.} All calculations are sampled on a 12 × 12 × 12 Gamma-centered grid of the BZ. Dashed lines show fits of Eq.~(\ref{eq:kappa_temperature}) to values calculated using the BTE-VDMFT approach.}
\label{figS4:kappaBTE}
\end{figure}

A comparison of thermal conductivities calculated using the GK-VDMFT, BTE-PT, and BTE-VDMFT approaches is illustrated in Fig.~\ref{figS4:kappaBTE}. The close consistency between the GK-VDMFT and BTE-VDMFT approaches highlights the importance of using accurate phonon lifetimes when employing the BTE-SMA approach.

Within the BTE-PT approach, the temperature dependence of the thermal conductivity arises from the heat capacity and the phonon lifetimes, and can thus be described using a simple expression,
\begin{equation}
    \kappa(T) \sim \frac{ca^2}{T^2}\frac{e^{a/T}}{(e^{a/T}-1)^2}\frac{e^{a/2T}-1}{e^{a/2T}+1}\,,
    \label{eq:kappa_temperature}
\end{equation}
where $c$ and $a$ are constants. This expression comes from the approximating the absorption lifetime given by the first term of Eq.~(\ref{eqn:linewidthPT}) under the assumption that $\Omega_{\lambda'}(\bm{k}') = \Omega_{\lambda''}(\bm{k}'') = \Omega_\lambda(\bm{k})$ and recovers the expected limits of the thermal conductivity: $\kappa\sim 1/T$ as $T\rightarrow\infty$, and $\kappa\sim e^{-1/T}/T^2$ as $T\rightarrow 0$. Figure~\ref{figS4:kappaBTE} shows fits of Eq.~(\ref{eq:kappa_temperature}) to the thermal conductivity values calculated using the BTE-VDMFT approach. While all fits show agreement with calculated values above 100\,K, they cannot simultaneously fit the calculated values across the full temperature range. In particular, while the fits show a trivial maximum of the thermal conductivity with temperature, they are not able to capture the narrow maximum of the thermal conductivity that is shown by the BTE-VDMFT calculations. This disagreement may be a signature of nonperturbative effects~\cite{ikeda_kondo-like_2019}, as discussed in the main text.

\section{Effects of grain boundary scattering on thermal conductivities}

At low temperatures, phonon lifetimes can be dominated by phonon-grain boundary scattering as opposed to phonon-phonon scattering. The phonon lifetimes calculated using SCP+PT or SCP+VDMFT, which only include the intrinsic effects of phonon-phonon scattering, can be adjusted using Matthiessen's rule to include the effects of phonon-boundary scattering:
\begin{equation}
    \tau^{-1}_{\lambda,\text{eff}}(\bm{k}) = \tau^{-1}_{\lambda}(\bm{k}) + 2v_\lambda(\bm{k})/L\,,
\end{equation}
where $v_\lambda(\bm{k})$ is the group velocity of mode $\lambda$, and $L$ is the grain size of the material.

\begin{figure}[ht]
\includegraphics[width=3.38in]{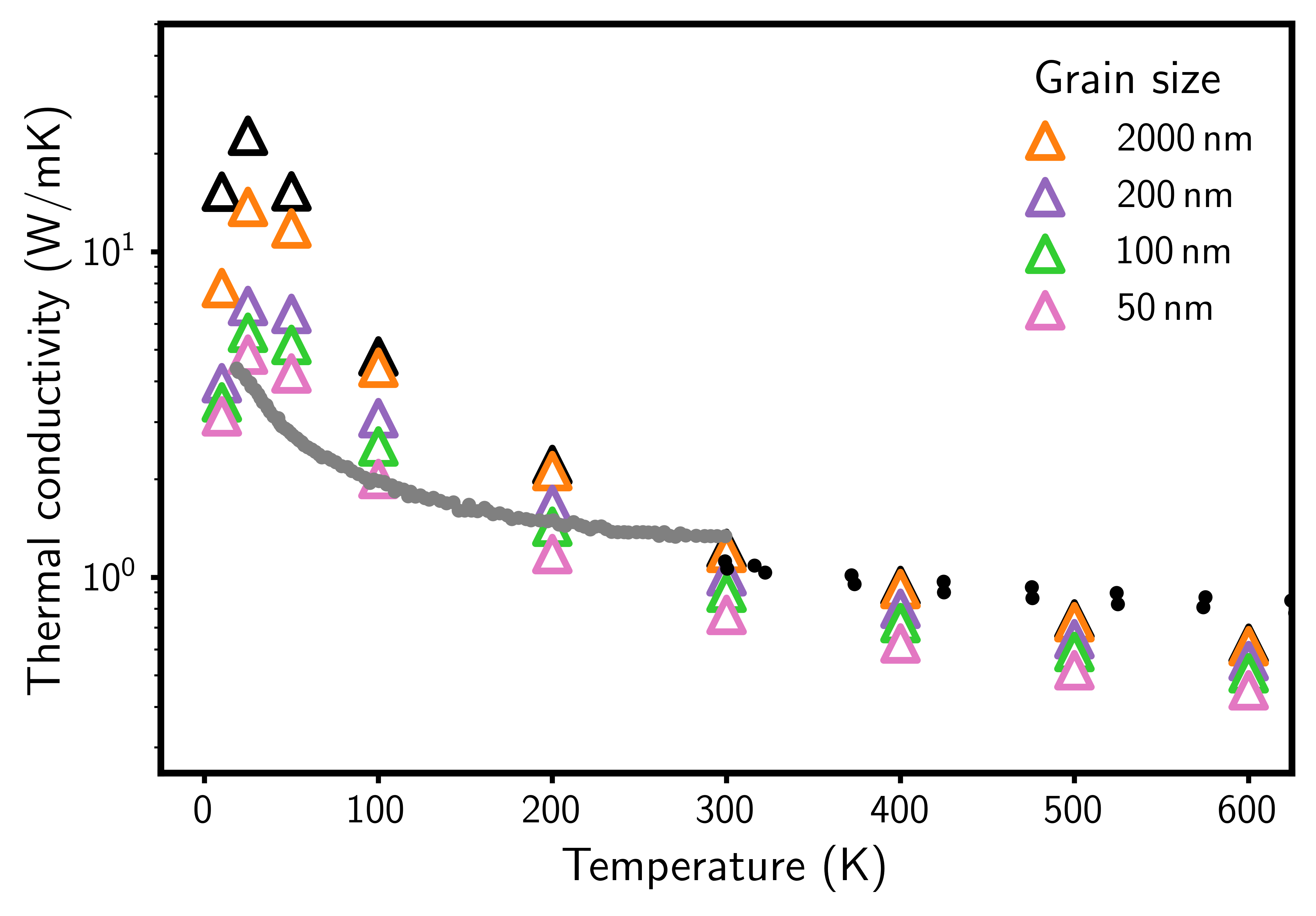}
\caption{Thermal conductivities of BaGG calculated using the BTE formalism [Eq. (1) of the main text] with lifetimes extracted from the SCP+VDMFT self-energy. The thermal conductivity calculated using intrinsic lifetimes are shown in the black symbols, while those calculated using effective lifetimes with different grain sizes are showed in the colored symbols. Grey and black points show experimental measurements of the thermal conductivity are from Refs.~ \citep{sales_structural_2001} and \citep{may_characterization_2009}, respectively.}
\label{figS5:grainBoundary}
\end{figure}

In Fig.~\ref{figS5:grainBoundary}, we show the BTE-VDMFT results for the thermal conductivity of BaGG including the effects of phonon-boundary scattering calculated using different grain sizes. A large grain size of $L=2\,\mu$m, used in Refs.~\onlinecite{tadano_impact_2015, tadano_quartic_2018}, has a negligible impact on the calculated thermal conductivities at temperatures greater than 100\,K. We find that a smaller grain size of $\sim$100\,nm needs to be used to find agreement between our calculated values and experimental measurements at temperatures less than 300\,K. However, using such a grain size also decreases the thermal conductivity at higher temperatures, such that they are lower than those measured experimentally.

%